\documentclass[epsfig,twocolumn,preprintnumbers,amsmath,amssymb]{revtex4}
\usepackage{epsfig}

\begin{document}

\title{
Energy dependence of elliptic flow over a large pseudorapidity
range in Au+Au collisions at RHIC}
\author {
B.B.Back$^1$, M.D.Baker$^2$, M.Ballintijn$^4$, D.S.Barton$^2$,
R.R.Betts$^6$, A.A.Bickley$^7$, R.Bindel$^7$, A.Budzanowski$^3$,
W.Busza$^4$, A.Carroll$^2$, Z.Chai$^2$, M.P.Decowski$^4$,
E.Garc\'{\i}a$^6$, T.Gburek$^3$, N.George$^{1,2}$,
K.Gulbrandsen$^4$, S.Gushue$^2$, C.Halliwell$^6$, J.Hamblen$^8$,
M.Hauer$^2$, G.A.Heintzelman$^2$, C.Henderson$^4$, D.J.Hofman$^6$,
R.S.Hollis$^6$, R.Ho\l y\'{n}ski$^3$, B.Holzman$^2$,
A.Iordanova$^6$, E.Johnson$^8$, J.L.Kane$^4$, J.Katzy$^{4,6}$,
N.Khan$^8$, W.Kucewicz$^6$, P.Kulinich$^4$, C.M.Kuo$^5$,
W.T.Lin$^5$, S.Manly$^8$, D.McLeod$^6$, A.C.Mignerey$^7$,
R.Nouicer$^2,6$, A.Olszewski$^3$, R.Pak$^2$, I.C.Park$^8$,
H.Pernegger$^4$, C.Reed$^4$, L.P.Remsberg$^2$, M.Reuter$^6$,
C.Roland$^4$, G.Roland$^4$, L.Rosenberg$^4$, J.Sagerer$^6$,
P.Sarin$^4$, P.Sawicki$^3$, H.Seals$^2$, I.Sedykh$^2$,
W.Skulski$^8$, C.E.Smith$^6$, M.A.Stankiewicz$^2$,
P.Steinberg$^2$, G.S.F.Stephans$^4$, A.Sukhanov$^2$,
J.-L.Tang$^5$, M.B.Tonjes$^7$, A.Trzupek$^3$, C.Vale$^4$,
G.J.van~Nieuwenhuizen$^4$, S.S.Vaurynovich$^4$, R.Verdier$^4$,
G.I.Veres$^4$, E.Wenger$^4$, F.L.H.Wolfs$^8$, B.Wosiek$^3$,
K.Wo\'{z}niak$^3$, A.H.Wuosmaa$^1$,
B.Wys\l ouch$^4$\\
$^1$ Physics Division, Argonne National Laboratory, Argonne, IL
60439-4843\\ $^2$ Chemistry and C-A Departments, Brookhaven
National Laboratory, Upton, NY 11973-5000\\ $^3$ Institute of
Nuclear Physics PAN, Krak\'{o}w, Poland\\ $^4$ Laboratory for
Nuclear Science, Massachusetts Institute of Technology, Cambridge,
MA 02139-4307\\ $^5$ Department of Physics, National Central
University, Chung-Li, Taiwan\\ $^6$ Department of Physics,
University of Illinois at Chicago, Chicago, IL 60607-7059\\ $^7$
Department of Chemistry and Biochemistry, University of Maryland,
College Park, MD 20742\\ $^8$ Department of Physics and Astronomy,
University of Rochester, Rochester, NY 14627\\ }
\date{\today}

\begin{abstract}\noindent
This paper describes the measurement of the energy dependence of
elliptic flow for charged particles in Au+Au collisions using the
PHOBOS detector at the Relativistic Heavy Ion Collider (RHIC).
Data taken at collision energies of $\sqrt{s_{_{NN}}} =$ 19.6,
62.4, 130 and 200 GeV are shown over a wide range in
pseudorapidity. These results, when plotted as a function of
$\eta'=|\eta|-y_{beam}$, scale with approximate linearity
throughout $\eta'$, implying no sharp changes in the dynamics of
particle production as a function of pseudorapidity or increasing
beam energy.

\end{abstract}

\maketitle


%%%% intro to flow, RHIC and SPS results %%%%%

The characterization of collective flow of produced particles by
their azimuthal anisotropy has proven to be one of the more
fruitful probes of the dynamics of heavy ion collisions at RHIC.
The elliptic flow signal ($v_{2}$) at midrapidity is significant
and consistent with expectations from hydrodynamic models at low
$p_{T}$ \cite{starpt}. It has been interpreted as evidence for the
production of a highly thermalized state, and perhaps for partonic
matter~\cite{gyulassy}. At high $p_{T}$, the observed shape of
elliptic flow~\cite{starjq1,starjq2} is consistent with
calculations incorporating jet quenching~\cite{JQ} and quark
coalescence~\cite{QC}. Interestingly, the fall of $v_{2}$ with
increasing pseudorapidity ($\eta$)~\cite{phflow} has been less
amenable to understanding~\cite{hirano}.

Given the wide range of pseudorapidity coverage and energies
available in PHOBOS data, it is interesting to examine the extent
to which the shape of the flow distributions change with energy in
the frame of reference of one of the incoming nuclei. The
multiplicity distribution has been examined by PHOBOS as a
function of $\eta'=|\eta|-y_{beam}$ (which is an approximation of
the rest frame of one of the nuclei) and found to be energy
independent over a wide range of $\eta'$~\cite{limfrag}. Data
showing such energy independence is said to be consistent with the
concept of ``limiting fragmentation"~\cite{limfragold}, which, as
used here, may extend well beyond the region of the collision
normally thought of as the fragmentation region. This work
examines the degree to which the elliptic flow in Au+Au collisions
at RHIC exhibits limiting fragmentation.

%%%% data used and detector as used in analysis %%%%%%%%%%

The PHOBOS detector employs silicon pad detectors to perform
tracking, vertex detection and multiplicity measurements. Details
of the setup and the layout of the silicon sensors can be found
elsewhere~\cite{phobos_det}. Detector components relevant for this
analysis include the first six layers of both silicon spectrometer
arms, the silicon vertex detector (VTX), the silicon octagonal
multiplicity detector (OCT), three annular silicon multiplicity
detectors on each side of the collision point, and two sets of
scintillating paddle counters.

Monte Carlo simulations of the detector performance were based on
the Hijing event generator~\cite{hijing} and the
GEANT~3.21~\cite{geant} simulation package, folding in the signal
response for scintillator counters and silicon sensors.

\begin{figure*}[t]
%\begin{figure}[h]
%\centerline{ \epsfig{file=V2eta_stacked_40_nokolb.eps,width=9.0cm}
\centerline{ \epsfig{file=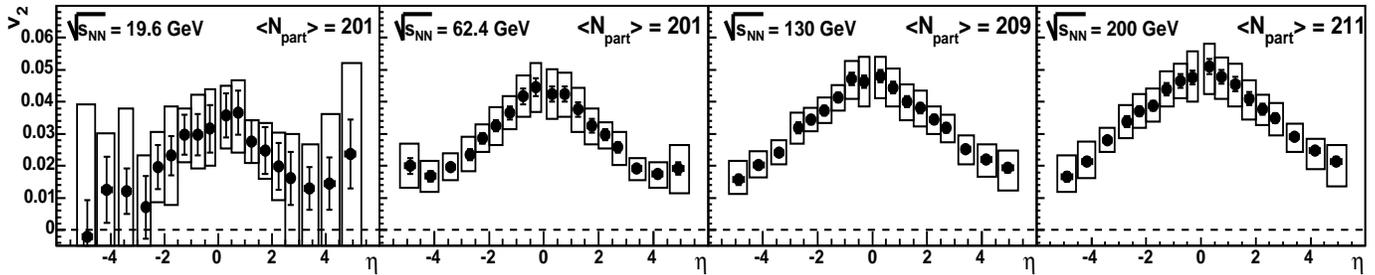,width=18.2cm,
height=4.0cm} } \caption{ The points represent the elliptic flow,
averaged over centrality (0-40\%), as a function of $\eta$ shown
separately for four beam energies. The boxes represent systematic
uncertainties at 90\% C.L. $\langle N_{part} \rangle$ gives the
average number of participants for each data sample.}
\label{v2stacked}
%\end{figure}
\end{figure*}

The data used in this analysis were recorded in the year 2000
($\sqrt{s_{_{NN}}}$=130 GeV), 2001 ($\sqrt{s_{_{NN}}}$=19.6 and
200 GeV) and 2004 ($\sqrt{s_{_{NN}}}$=62.4 GeV)  runs. Details on
the event selection and signal processing can be found elsewhere
\cite{limfrag,phflow}.  The majority of the data was taken with
zero magnetic field to simplify the analysis.  The exception was
at $\sqrt{s_{_{NN}}} =$130 GeV, where field-on data was included
to increase statistics. The selected data at each of the four
energies correspond to the 40\% most central events.

The analysis presented here is very similar to that used in
previously published PHOBOS results at $\sqrt{s_{_{NN}}}$=130 GeV
\cite{phflow}. It is based on the anisotropy of the azimuthal
distribution of charged particles traversing the detector. At the
points where charged tracks pass through an active silicon
detector, energy is deposited in the form of ionization. The pad
where energy is deposited is said to be a ``hit".  This analysis
is based on the ``subevent" technique where one studies the
correlation of hits in one part of the detector with the event
plane angle as determined by hits in a different part of the
detector~\cite{pandv}.

The strength of the flow is given by the $n^{ th}$ Fourier
coefficient of the particle azimuthal angle distribution,
\begin{equation}
 \frac{dN}{d(\phi-\psi_{R})}
\sim 1 + \sum_{n}2{\rm v}_{n}\cos[n(\phi-\psi_{R})],
\end{equation}
where $\psi_{R}$ is the true reaction plane angle defined by the
impact parameter and the beam axis. This analysis was confined to
$n=2$, the so-called elliptic flow.  $v_{2}$ and $\psi_{2}$ (our
best estimate of $\psi_{R}$) are calculated as in
reference~\cite{pandv}.

The analysis presented here differs from that in our previously
published elliptic flow results~\cite{phflow} by including only
collisions within $\pm$10 cm of the nominal vertex position (along
the beam axis). This constraint encompasses the bulk of our data
at all energies and satisfies centrality determination restraints
at 19.6 GeV.
%A significant difference between this analysis and that in
%reference~\cite{phflow} is that this work is based on data from
%collisions within $\pm$10 cm of the nominal collision point along
%the beam axis. The need to use data with collision points in this
%region of the detector is driven by the fact that it constitutes
%the bulk of our data at all energies, and the entirety of the data
%at 19.6 GeV.
%
Unfortunately, such collisions occur in a region of the detector
that is quite non-uniform in azimuth.
%, as can be seen in
%Figure~\ref{detmap}, which schematically depicts the `unrolled'
%silicon in the OCT, RP and RN detectors used for this analysis.
Above and below and to each side of the nominal collision point
are holes in the OCT subdetector amounting to half the azimuthal
coverage. Particles passing into the region above or below the
collision point traverse the VTX detector. The holes to each side
of the nominal collision point prevent shadowing of the
spectrometer detectors.

\begin{figure}[h]
\centerline{ \epsfig{file=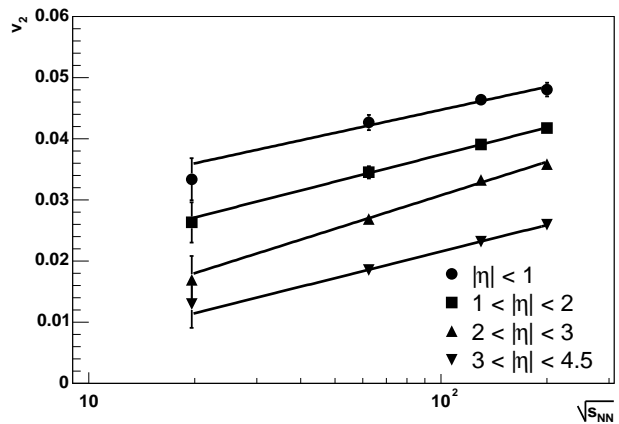,width=9.0cm} }
\caption{ The evolution of the
 elliptic flow in $\eta$ bins, averaged over
 centrality (0-40\%), as a function of $\sqrt{s_{_{NN}}}$.  The
 lines represent fits to the data.
 } \label{v2roots}
\end{figure}

The detector hit map was symmetrized in order to use the subevent
technique for the flow analysis.  For each of the holes above and
below the collision point, this entailed filling the hole by
mapping the inner VTX layer hits onto a virtual OCT layer. Because
of the limited coverage of the silicon spectrometer, the holes to
each side were filled by linearly extrapolating across the hole
the hit density adjacent to the hole region in $\phi$ on an
event-by-event basis. The latter procedure yielded an expected
loss in sensitivity, reducing the raw, measured flow signal by
roughly 10\% relative to a detector with no hole in that region.

Before analyzing each event for flow, the symmetrized detector hit
map was weighted to correct for the relative phase space
differences between detector pads due to geometry (acceptance
weight), and the dilution of the asymmetries in the hit map due to
the occupancy in the pixelized detector (occupancy weight).

The acceptance weights were calculated in each $\eta$ annulus
through the use of individual hit weights, w$_{i}^{a}$, which are
proportional to the inverse of the average number of hits in each
pad.  These weights were determined separately in bins of
centrality and longitudinal vertex position.

The occupancy was determined on an event-by-event basis from the
number of occupied ($N_{\rm occ}$) and unoccupied ($N_{unocc}$)
pads in small sections of the detector. The occupancy weight in a
given section, representing the average number of tracks per hit
pad, was determined assuming a Poisson statistical distribution as
\cite{multdists}
\begin{equation}
{\rm Occ}(\eta,\phi)=\frac{\mu}{1-e^{-\mu}},
\end{equation}
where $\mu$=$\ln$(1 + $N_{occ}$/$N_{unocc}$) is the average number
of tracks per pad. This occupancy was used in concert with the
acceptance weight to produce the overall weight for a given hit,
\begin{equation}
{\rm w}_{i}={\rm w}^{a}_{i}{\rm Occ}(\eta_{i},\phi_{i}),
\end{equation}
which was used in the determination of $\psi_{2}$.

Using the weighted and symmetrized hit map, the
resolution-corrected elliptic flow was calculated with the
standard subevent technique used for our earlier
results~\cite{phflow}. The subevent regions used in the event
plane calculation were $0.1<|\eta|<3.0$ for all four energies. The
event plane resolution was calculated separately for each
centrality bin. The resolution correction ranged from 2 to 3 on
average, with the larger correction necessary at 19.6 GeV. For the
determination of $v_{2}$ in the positive (negative) $\eta$ region
of the detector, the subevent on the opposite side of midrapidity
was used to evaluate $\psi_{2}$.

\begin{figure}[h]
\centerline{ \epsfig{file=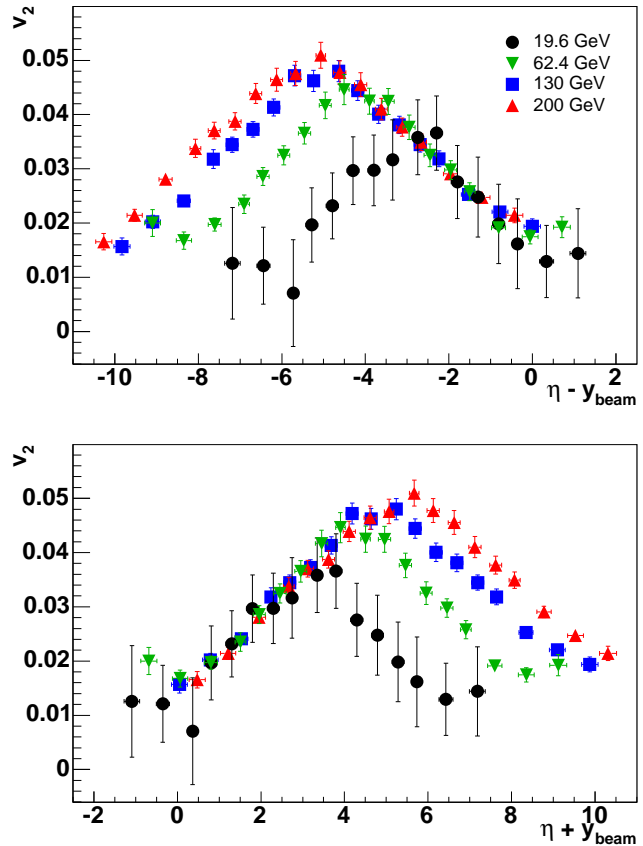,width=9.0cm}
} \caption{ The
 elliptic flow, averaged over
 centrality (0-40\%), as a function of $\eta-y_{beam}$ (top)
 and $\eta+y_{beam}$ (bottom) for each of the four
energies studied in this paper. The error bars represent the
1$\sigma$ statistical errors only.} \label{v2plusminusetaprime}
\end{figure}

Monte Carlo simulations showed a residual suppression of the flow
signal, thought to be dominated by background particles carrying
no flow information and the loss of sensitivity due to the hit map
symmetrization and the occupancy correction algorithm. As in our
earlier work, this suppression was corrected using simulated data
by comparing the output resolution corrected flow signal to the
input flow signal for many samples of simulated data with
different shapes and magnitudes of input flow.

%%%%%%%%%%%%%% systematic errors %%%%%%%%%%%%%%%%%%%%%%

 Numerous sources of systematic error were investigated, including
effects due to the hit definition, hit merging, subevent
definition, knowledge of the beam orbit relative to the detector,
shape of the dN/d$\eta$ distribution, hole filling procedure,
vertexing algorithm, transverse vertex cuts, magnetic field
configuration and suppression correction determination. The effect
of these sources depended both on $\eta$ and centrality. In
general, the systematic error arising from each source was
determined by varying that specific aspect of the analysis (or
several aspects in concert) within reasonable limits and
quantifying the change in the final $v_{2}$ result as a function
of $\eta$ and centrality. The individual contributions were added
in quadrature to derive the 90\% confidence level error shown in
the results presented here.  The systematic uncertainty was
dominated by the suppression correction determination.

%%%%%%%%%%%%% results %%%%%%%%%%%%%%%%%%%%%%%%%%%%%%%

The fully corrected elliptic flow signal is shown for all four
energies in Figure~\ref{v2stacked}. The values shown are
consistent with previous measurements where there is energy and
acceptance overlap~\cite{phflow,na49,starcumulant}. The error bars
represent the 1$\sigma$ statistical errors and the boxes give a
measure of the systematic error for each point at 90\% confidence
level.  The statistical errors are somewhat correlated
point-to-point due to shared event plane and event plane
resolution determinations.

Relative to the other energies, the data at 19.6 GeV comprise a
smaller set of events with both smaller flow and multiplicity.
This leads to the lack of statistical power at 19.6 GeV apparent
in Figure~\ref{v2stacked}. This, in turn, contributes to the large
systematic errors because of the difficulty in separating
statistical and systematic effects.

%For ease of comparison, the curves are shown on the same plot in
%Figure~\ref{v2eta},with systematic errors removed for clarity.

The PHOBOS 200 GeV $p_{T}$-integrated track-based results agree
very well with the data shown in Figure~\ref{v2stacked} in the
available range of $0<\eta<2.0$~\cite{qm04}. Also, the PHOBOS
track-based elliptic flow results are consistent with the STAR
4-particle cumulant results as a function of
$p_{T}$~\cite{qm04,starcumulant}. This agreement, along with the
fact that the track-based technique is expected to have a
different (and smaller) susceptibility to non-flow correlations,
implies the hit-based results shown here do not have a significant
contribution from non-flow correlations, at least in the region
$|\eta|<$2.

All four energies in Figure~\ref{v2stacked} show a
non-boost-invariant, roughly triangular shape peaking at
midrapidity. At the lower energies the flow seems to level off
(and maybe even rise) at high $|\eta|$.  This might be due to
pronounced directed flow in these regions at the lower energies or
an effect due to the presence of participant nucleons. At higher
energies, the participants are pushed further out in $|\eta|$ and
the directed flow is smaller~\cite{qm04}.

Figure~\ref{v2roots} shows that the magnitude of the elliptic flow
grows logarithmically with the beam energy for differing regions
of $|\eta|$.  The lines represent fits to the data.

%The growth is linear for
%differing regions of $|\eta|$. Since the evolution of $v_{2}$ with
%$|\eta|$ is not exactly linear, the slopes vary somewhat as a
%function of the $|\eta|$ region.

%
\begin{figure}[h]
\centerline{
\epsfig{file=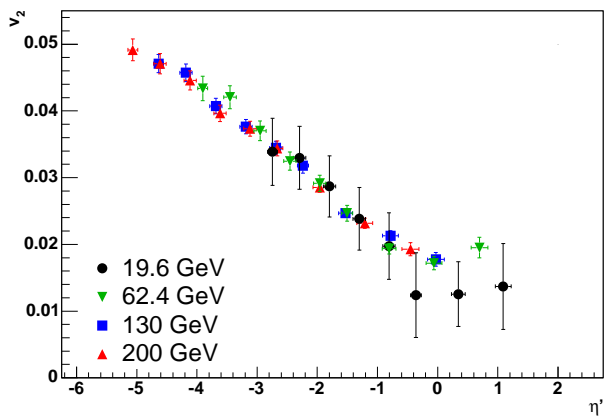,width=9.0cm} }
 \caption{
Elliptic flow, averaged over centrality (0-40\%), as a function of
$\eta'=|\eta|-y_{beam}$ for four beam energies. The error bars
represent the 1$\sigma$ statistical errors only.}
\label{v2etaprime}
\end{figure}

Figure~\ref{v2plusminusetaprime} shows the results from
Figure~\ref{v2stacked} plotted in terms of $\eta \pm y_{beam}$.
For clarity, only the statistical errors are shown.  For this and
the following plot, the highest $|\eta|$ points in the 19.6 GeV
data are not included because they lack significance due to large
systematic errors.

For the purpose of examining boost invariance and the limiting
fragmentation behavior of the elliptic flow, it would be best to
use the rapidity ($y$) rather than $\eta$. Unfortunately, this is
not possible with the PHOBOS detector over such a large acceptance
range. The effect of working in $\eta$ rather than $y$ is
estimated by us and others~\cite{kolb} to be small ($<$15\%) and
is not thought to change the qualitative features of the shapes,
though deviations near midrapidity might occur if plotted as a
function of $y \pm y_{beam}$.

%However, the behaviors of the $p_{T}$-integrated multiplicity and
%elliptic flow under transformations between $\eta$ and y have been
%estimated by us and others~\cite{kolb}. The most significant
%effect when going from y to $\eta$ occurs near midrapidity where
%the multiplicity dips and the elliptic flow peaks.  This effect is
%estimated to be approximately 10\% near midrapidity, with the
%difference falling rapidly to zero near $|\eta|=2$.  Our estimates
%agree with those shown in reference~\cite{kolb}. The effect is
%small and does not change the qualitative features of the curves.

Figure~\ref{v2etaprime} shows the elliptic flow, as seen in
 Figure~\ref{v2stacked}, where data from positive and negative
 $\eta$ are averaged and plotted as a function of
$\eta'=|\eta|-y_{beam}$ for all four energies. The four curves
scale throughout the region of $\eta'$ overlap through mid-$\eta$
for each energy.  This scaling, along with the fact that the shape
in $\eta'$ is approximately linear, implies the triangular shape
of $v_{2}$($\eta$) in Figure~\ref{v2stacked}
 and the linear evolution of $v_{2}$($\ln\sqrt{s_{_{NN}}}$) shown
in Figure~\ref{v2roots}.

%%%%%%%%%%%%%%%%%%% conclusions %%%%%%%%%%%%%%%%%%%%%%%%%%%%%

The results in Figure~\ref{v2etaprime} show that elliptic flow
exhibits limiting fragmentation in the full range of $\eta'$,
reminiscent of what was observed in the
multiplicity~\cite{limfrag}. The degree to which the elliptic flow
is shown to be independent of energy everywhere in $\eta'$ is
somewhat surprising given the success of hydrodynamics in
describing the flow at the higher energies in the region near
mid-$\eta$.

In summary, these results
%presented here
%demonstrate the validity of the concept of limiting fragmentation
%for
illustrate the energy-independence of elliptic flow  in
ultra-relativistic heavy ion collisions over a large region of
$\eta'$ throughout the energy reach of RHIC. The degree to which
the energy independence of the results extends to midrapidity for
the elliptic flow is intriguing. It is difficult to reconcile this
fact with the common assumption that the particle production at
midrapidity differs from that in the fragmentation region,
particularly at the higher energies. These results are not
obviously compatible with the underlying (Bjorken/Feynman) picture
of the nuclear collisions at RHIC, supported by the success of
hydrodynamics at midrapidity. They imply the longitudinal degree
of freedom is not to be treated trivially in our experimental and
theoretical efforts to understand these collisions.

Acknowledgements:
%We acknowledge the generous support of the
%Collider-Accelerator Department
% (including RHIC project personnel) and Chemistry Departments at BNL.  We
% thank Fermilab and CERN for help in silicon detector assembly.  We thank the
% MIT School of Science and LNS for financial support.
%
This work was partially supported by U.S. DOE grants
DE-AC02-98CH10886, DE-FG02-93ER40802,
DE-FC02-94ER40818,  % MIT
DE-FG02-94ER40865, DE-FG02-99ER41099, and W-31-109-ENG-38, by U.S.
NSF grants 9603486, % Phobos TOF
0072204,            % Rochester until 6/03
and 0245011,        % Rochester starting 6/03
by Polish KBN grant 2-P03B-10323, and by NSC of Taiwan Contract
NSC 89-2112-M-008-024.

\end{document}